\magnification=1200
\nopagenumbers
\vglue 2.0truein
\centerline {\bf CAN A UNRUH DETECTOR FEEL A COSMIC STRING?}
\vskip .3truein
\vskip .9truein
\centerline { A.H.Bilge $^{\dagger}$, M.Horta\c csu $^{\dagger,*}$,
N. \" Ozdemir $^{*}$ }
\vskip .4truein
\centerline {$^ \dagger$ TUBITAK Marmara Research Center, Gebze, Turkey}
\vskip .3truein
\centerline{$^{*}$  Physics Dept., I.T.U., Maslak, Istanbul, Turkey}
\vskip 1.3truein
\noindent{Abstract.} Unruh's detector calculation is used to study
the effect of the defect angle $\beta$ in a space-time with a cosmic string
for both the excitation and deexcitation cases.
It is found that a rotating detector results in a non-zero effect for both
finite (small) and infinite (large) time.
\vskip .3truein
\vfill\eject
\baselineskip=18pt
\footline={\centerline{\folio}}
\pageno=1
\noindent
{\bf I. INTRODUCTION}
\vskip 0.3cm

Different aspects of cosmic strings are studied in many papers and reviews.
One can give References 1,2,3 as a good point to start learning about this
ever developing field. Among new physical processes where the effects of
cosmic strings are studied one can cite references 4 and 5. Here
stimulated and spontaneous emission near cosmic strings are studied.
The presence of the cosmic string gives rise to modifications in the rates
of these processes.

Here we do the similar calculation as in these references  for different
physical processes,
 using the model of a particle detection due to
Unruh $^{/6}$ and De Witt $^{/7}$.

Section II is devoted to the review of the method and the results already
known.
We first go  over the Davies-Sahni $^{/8}$  results for the detector at rest
and
oscillating in the $r$ and $z$ directions. Note that   if  the detector
switches on for a
finite time $T$, the  response function depends on $T$, the  excitation energy
$E$,  and the   distance from the  string $R$ $^{/9}$. Here we will stick to
the the standard method of  $^{/10}$,  performing the calculation for
infinite time.

In Section III we study the case when the detector revolves around
the string at distance $R$ with
constant angular velocity
$\omega$. This case was also studied by Davies and Sahni $^{/8}$, with no
definite result.  We do the computation both for finite (small) and infinite
(large) time and we find a change in the detector response function for both
cases. For the deexcitation
amplitude, we find extra poles if the string parameter $\beta$ is less
than a definite value. We conclude with a few remarks.

\bigskip
\noindent
{\bf II.  REVIEW OF PREVIOUS RESULTS}

As  described in references 6,7, and 10, here we assume that an idealized
point particle, acting as a detector with internal energy levels labelled by
the energy $E$ is coupled via a monopole interaction with a scalar field
$\phi$.  The particle detector moves along a world line described by the
function $x^{\mu} (\tau) $, where $\tau $ is the detector's proper time.
The detector-field interaction is described by an interaction Lagrangian
$ g\  m(\tau)\  \phi (x(\tau)) $, where $g$ is a small coupling constant
and $m(\tau)$ is the detector's monopole moment operator.

The calculation is performed in first order perturbation theory. We
square the first order amplitude and sum over all the energies and
scalar field excited states to get the transition probability
$$ g^{2} \sum _{E} |\langle E|m(0)|E_0\rangle |^{2} F(E-E_0) \eqno {(1)} $$
where  the detector
response function is given by
${\it F}(E-E_0)$
$${\it F}(E-E_0)={\int_{-\infty}^\infty} d\tau
{\int_{-\infty}^\infty}d\tau^{\prime}
\exp{\left[-i(E-E_0)(\tau-\tau^{\prime})\right]}
G^+(x(\tau),x(\tau^{\prime})). \eqno{(2)}$$
Here $G^+$ is the Wightman function of a scalar particle for the metric in
question; $\tau$ is the proper time and $E-E_0>0$, is the excitation of the
detector. If we consider the transition probability per unit time, we have
to consider
$$g^2\sum_E \vert\langle E\vert m(0)\vert E_0\rangle \vert^2
\int_{-\infty}^\infty
d(\Delta\tau) e^{-i(E-E_0)\Delta\tau} G^+(\Delta\tau) \eqno{(3)}$$
where $\Delta\tau=\tau-\tau^{\prime}$.

We ascribe a certain trajectory to the detector and  look for possible
non-zero response.  The effect depends not only on the metric, but also
on the worldline followed by
the detector.  If we get zero response for a certain trajectory, this does
not at all mean that we will get zero response for other trajectories for
the same metric. In fact quite the contrary is known to be true.  It is well
known that when the detector accelerates in Minkowski space, we get a nonzero
response, the Unruh effect, whereas when the detector is stationary, or
moving with constant velocity, we get null result in the same space.
On the other hand, in de Sitter space and for other non-flat metrics or
in the presence of thermal radiation $^{/8,10}$, even a stationary trajectory
gives a non zero result.

Note that the results above refer to a detector calculation for
infinite time. As a limiting case of the detector response for
finite time, in Section 3.2  we study the behaviour of the integrand near
$\Delta \tau=0$. We show that the ``pole" as
$(\Delta \tau-i\epsilon)^{-2}\to 0$ depends on
$\beta\omega R$ only, hence a ``background contribution" can be subtructed
to regularize the integral. Then the first nonzero term in the Laurent
expansion
of the regularized integrand will be nonzero for $\beta\ne {1\over k}$,
$k$ integer, and zero otherwise.

Here we will study the response function in the cosmic string background.
We anticipate variation from the Minkowski result if the presence of the
cosmic string actually changes the physics of the problem. We introduce the
Wightman function
to this formalism, already calculated by many authors
$^{/11}$. We use the form given by the expression
$$G^+={1\over{(2\pi)^2\beta r_1r_2}}\left[
{1-\Delta^{2\over\beta}\over{1+\Delta^{2\over\beta}-2\Delta^{1\over\beta}
\cos(\phi-\phi^{\prime})}}\right] \eqno{(4)}$$
where
$$r_1=[-(t-t^{\prime}-i\epsilon)^2+(z-z^{\prime})^2+(r-r^{\prime})^2]
^{1\over 2}, \eqno{(5)}$$
$$r_2=[-(t-t^{\prime}-i\epsilon)^2+(z-z^{\prime})^2+(r+r^{\prime})^2]
^{1\over 2}, \eqno{(6)}$$
$$\Delta=\left({r_2-r_1\over{r_2+r_1}}\right), \eqno{(7)}$$
for the metric
$$ds^2=dt^2-dr^2-dz^2-\beta^2 r^2 d\phi^2 . \eqno{(8)}$$
Here $\beta $ is a constant satisfying $ 0< \beta\leq 1$.

In calculating $F(E)$, we first study a detector at rest. It is shown that
at the coincidence limit for $z, r$ and $\phi$,
the response function $F(E)$ is per unit time proportional to
$${1\over{(2\pi)^2}}\int_{-\infty}^\infty  d(\Delta t)
{e^{-i(E-E_0)(t-t^{\prime})}\over{\beta(-i(t-t^{\prime}-i\epsilon))
\sqrt{4r^2-(t-t^{\prime}-i\epsilon)^2}}}$$
$$\times{
\{\sqrt{4r^2-(t-t^{\prime}-i\epsilon)^2}+i(t-t^{\prime}-i\epsilon)\}
^{1\over\beta}+
\{\sqrt{4r^2-(t-t^{\prime}-i\epsilon)^2}-i(t-t^{\prime}-i\epsilon)\}
^{1\over\beta}
\over{
\{\sqrt{4r^2-(t-t^{\prime}-i\epsilon)^2}+i(t-t^{\prime}-i\epsilon)^2\}
^{1\over\beta}-
\{\sqrt{4r^2-(t-t^{\prime}-i\epsilon)^2}-i(t-t^{\prime}-i\epsilon)^2\}
^{1\over\beta}}}. \eqno{(9)}$$
As noted in reference 6  this expression has poles at $(t-t^{\prime})=
i\epsilon$. There are no cuts in the lower half plane.
If we close the contour in the lower half plane, as we should since
$E-E_0>0$ we get zero, the Minkowski result.
Reference 9   shows that performing a finite integral in proper time gives
non zero results.

To study the accelerating case we first note that  in this space the
equations of motion are
$$\eqalignno{
\ddot x^{\phi}+{1\over r}\dot x^{\phi}\dot x^{r}&=F^\phi ,&(10)\cr
\ddot x^{r}-\beta^2 r\dot x^{\phi}\dot x^{\phi}&=F^r,&(11) \cr
\ddot x^{t}&=F^t,&(12)\cr
\ddot x^{z}&=F^z.&(13)\cr}$$
If the force is harmonic, i.e.,
$$F^t={t\over{\alpha^2}},\quad F^r={r\over{\alpha^2}},\quad F^\phi=0,\quad
F^z=0,\eqno(14)
 $$
then we obtain
$$t=\alpha\sinh{\tau\over\alpha},\quad \quad
r=\alpha\cosh{\tau\over\alpha} .\eqno(15)$$
Then the response function reads
$${F(E)\over T}={1\over{\beta(2\pi)^2}}\int_{-\infty}^\infty
{d(\Delta\tau)e^{-i(E-E_0)\Delta\tau}\over
{4\alpha^2\sinh({\Delta\tau\over{2\alpha}}-i\epsilon)
\cosh({\tau+\tau^{\prime}\over
{2\alpha}})}}{A^{1\over\beta}+B^{1\over\beta}\over
{A^{1\over\beta}-B^{1\over\beta}}} \eqno(16)$$
where
$$A=\cosh{\tau+\tau^{\prime}\over{2\alpha}}+
i\sinh\left({\tau-\tau^{\prime}\over{2\alpha}}-i\epsilon\right)\eqno(17) $$
$$B=\cosh{\tau+\tau^{\prime}\over{2\alpha}}-
i\sinh\left({\tau-\tau^{\prime}\over{2\alpha}}-i\epsilon\right).\eqno(18) $$
In these expressions $-i\epsilon$ are put inside the hyperbolic
cosine and sine functions.  This is only correct when $\tau$ approaches
$\tau'$, the only point where $\epsilon$ has any meaning.

Since $\epsilon>0$, we do not get any cuts in the lower half plane.

We can perform the contour integration. The poles are at points where
$\sinh({\Delta\tau\over{2\alpha}}-i\epsilon)$ vanish.
Expanding the expression about the poles result in ,
$(T={\tau+\tau^{\prime}\over 2})$
$${F(E)\over T}={1\over{(2\pi)^2\beta}}
\int_{-\infty}^\infty
{d(\Delta\tau)\over{\cosh{T\over\alpha}}}e^{-i(E-E_0)\Delta \tau}
{(\cosh{T\over\alpha})^{1\over\beta}
[2+{{\sinh^2{\Delta\tau\over\alpha}-i\epsilon}\over
{\cosh^2{T\over\alpha}}}{1\over\beta}({1\over\beta}-1)+\cdots]
\over{
[\cosh({T\over\alpha})^{1\over\beta}{2\over\beta}
{\sinh^2
({\Delta\tau\over{2\alpha}}-i\epsilon)]\over{
\cosh{T\over\alpha}}}}}. \eqno{(19)}$$
All reference to $\beta$ nicely cancel. We end up with
$${F(E)\over T}={1\over{(2\pi)^2}}\int_{-\infty}^\infty
{d(\Delta\tau)\over{\alpha^2}}
{e^{-i(E-E_0)(\Delta\tau)}\over{\sinh^2{\Delta\tau\over{2\alpha}}
-i\epsilon}}={1\over{2\pi}}
{E-E_0\over{
[e^{2\pi(E-E_0)\alpha}-1]}} \eqno{(20)}$$
which is the Minkowiski result for an accelerating detector, or for a
particle in a heat bath with temperature
$T={1\over{2\pi\alpha k_B}}$, here $k_B$, is the Boltzman constant$^{/8}$.

We can also accelerate our detector parallel to the string.
The above result does not change if we take
$z=\alpha\cosh{\tau\over{\alpha}},t=\alpha\sinh{\tau\over\alpha}$
which corresponds to taking
$F_r=0, F_\phi=0, F_t={t\over{\alpha^2}}, F_z={z\over{\alpha^2}}$
When we set $\phi=\phi^{\prime}, r=r^{\prime}$, we get the same result,
namely
$${F(E-E_0)\over T}={1\over{2\pi}}{{E-E_0}\over{e^{2\pi(E-E_0)\alpha}-1}}.
\eqno{(21)}$$
Note that moving paralel or perpendicular to the string does not matter.

These results are true for the infinite contour when we do not impose a
cut-off on the interaction.  If the interaction is switched off after a
certain time, one finds that a finite effect due to the presence of the
string is detected as shown in reference 9.

We only review the results prior to the publication of reference 9 in this
section and refer to the original article for the situation for finite time
in the stationary string case.
We will, however, treat a new case with finite contour in the next section
and show that there exists, indeed, a finite  effect for this example.

For all these cases the integrand near $\Delta \tau=0$ is studied and there
is a nonzero qualitative effect. The case $\beta=1/k$ is indistinguishable
from $\beta=1$ provided that certain other parameters are kept constant.

\bigskip
\noindent
{\bf III. ROTATING DETECTOR}

We consider now a detector  rotating around the cosmic string  in a plane
perpendicular to the $z$ axis. In this case
$$F^\phi=F^t=F^z=0,\quad F^r=-A_\beta^2 \beta^2 \omega^2 R,$$
with
$$A_\beta={A\over \sqrt{1-\omega^2\beta^2 R^2}}.$$
Here $A$ is a constant and $R$ is the distance from the cosmic string. It
can be seen that then the trajectory is given by
$x^r={\rm const.}$, $x^t={\rm const.}$,  $x^z={\rm const.}$ and
$x^\phi=\omega t$, with $t= A_\beta \tau$.

We take $z=z^{\prime}$, $r=r^{\prime} = R $,
$\phi-\phi'=\omega(t-t^\prime)$. Then
$$\eqalignno{
 r_1&=i(t-t^\prime -i\epsilon),\cr
 r_2&=\sqrt{4R^2-(t-t^\prime-i\epsilon)^2},\cr
\cos(\phi-\phi')&=\cos\omega(t-t^\prime).\cr}$$
Writing $\Delta={\Delta_-\over\Delta_+}$ with
$\Delta_\pm= r_2\pm r_1,\qquad$
$F(E-E_0)/ T$ reduces to
\vfill\eject
$$\eqalignno{{F(E-E_0)\over T}&={1\over{(2\pi)^2\beta}}\cr
\times&\int_{-\infty}^\infty
{d(\Delta(\tau))e^{-i(E-E_0)\Delta\tau}\over
{i(\Delta t -i\epsilon)\sqrt{4R^2-(\Delta t-i\epsilon)^2}}}
{(\Delta_+^{1\over\beta}+\Delta_-^{1\over\beta})
(\Delta_+^{1\over\beta}-\Delta_-^{1\over\beta})
\over{
\Delta_+^{2\over\beta}+\Delta_-^{2\over\beta}
-2\Delta_+^{1\over\beta}\Delta_-^{1\over\beta}
\cos\omega(t-t^{\prime} )}}&(22)\cr}$$
Note that for finite $\epsilon$, the integrand is technically non-divergent,
However for small $\epsilon$ the above integral cannot be computed
numerically. In Section 3.1 we shall introduce a coordinate
transformation to convert the integral over the real line  to a contour
integral in the complex plane and use  residue calculus to obtain the result.
In Section 3.2, we shall study the behaviour of the integrand near $\Delta
\tau=0$ and we will use a Laurent expansion of the integrand around the
``pole" to study the divergence.

\noindent
{\bf 3.1 Infinite (large) time behaviour.}

To simplify the integral (22),  we make the change of variable
$$t-t^\prime -i\epsilon\to  2R \sin(z)$$
 where $z$ is the complex variable,
$z=x+iy$. Then $r_2=\pm 2R\cos z$, however it can be seen that  the
integrand is independent of the sign of $r_2$. Taking the positive sign,
the new parametrization gives
$$ \Delta_{+} = 2R e^{iz},\quad \quad
   \Delta_{-} = 2R e^{-iz}.$$
Using also $\Delta \tau=\Delta t/A_\beta$,
 our integral simplifies to
$$ I={1\over{R(2\pi)^{2} \beta A_{\beta}}} \int dz
e^{-i{{(E-E_{0})}\over {A_{\beta}}}\left(2R\sin(z) +i\epsilon \right)}
{{ \sin {z \over {\beta}} cos {z \over {\beta}}} \over {\sin z \left(
cos{2z\over{\beta}}-cos(2 \omega R (\sin z +i\epsilon))\right)}} \eqno
{(23)}$$

\medskip
\noindent
{\bf The case $E-E_0>0$.}
\smallskip

We first consider the case $E-E_0 >0,$ hence we close the contour in the
lower half plane. The imaginary part of the expression
$$ \Delta t -i\epsilon = 2R(\sin(x) \cosh (y)+i\cos(x) \sinh(y)).
\eqno{(24)}$$
defines the contour of integration and its real parts determine the lines of
constant $\Delta\tau$.
It can be seen that the integrand vanishes as $y\to -\infty$, hence the
contour integral can be evaluated using residues.
For small $\epsilon$, the product $\cos(x)\sinh(y)$ is small hence $x\to
\pi/2$ as $y\to -\infty$. Thus the contour of integration looks like the
union of straight lines $\{x=\pm \pi/2\}, y<0$ joined by a curve just below
the $x$-axis.

The poles inside the contour are the zeros of
$cos{2z\over \beta}-cos(2 \omega R (\sin z +i\epsilon))$. Using the identity
$\cos p-\cos q=-2 \sin({p+q\over 2})\ \sin({p-q\over 2})$ with $p={2z\over
\beta}$, $q=2\omega R(\sin z+i\epsilon)$, it can be seen that the
zeros of poles correspond to
$${x\over \beta} \pm \omega R \sin x\ \cosh y=-k\pi,\quad\quad
  {y\over \beta} \pm \omega R \cos x\ \sinh y\pm \epsilon/2=0. $$
However, as for  $y$ and  $\sinh y$ have the same sign, for
small $\epsilon$ the second equation can be satisfied only with the
negative sign.  Furthermore it can also be seen that $k$ has to be positive.

In the formulation $\epsilon $ was introduced to avoid the
poles at on the real axis. Hence after restricting the poles to the ones that occur for
$\epsilon >0$ we can take the limit $\epsilon \to 0$ and the only poles are
now given by $$x_k-\beta \omega R \sin x_k\ \cosh y_k=-\beta k\pi,\quad k>0,
\quad\quad   y_k- \beta \omega R \cos x_k\ \sinh y_k=0. \eqno (25) $$
It can also be seen that as $x\to -x$ the integrand for $z=x+iy$ goes to
negative of its complex conjugate, hence the contour integral is real.
The integral around the contour can be evaluted using residues as
$$I=-{1\over 4}
{2\pi \ i
\over R(2\pi)^2 \beta A_\beta}
\sum_{k=-\infty}^{k=+\infty}
{e^{-i{(E-E_0)\over A_\beta}(2R\sin z_k)}
  \sin {2z_k \over {\beta} } \over
{\sin z_k \
\sin ({z_k\over \beta}+\omega R\sin z_k)\
\cos ({z_k\over \beta}-\omega R\sin z_k)\
     ( {1\over \beta}-\omega R\cos z_k) } }$$
$$\eqno {(26)}$$
At the poles
$\sin ({z_k\over \beta}-\omega R\sin z_k)=0$ hence
$\cos ({z_k\over \beta}-\omega R\sin z_k)=(-1)^k$ and it can be seen that
$\sin ({z_k\over \beta}+\omega R\sin z_k)=\sin {2z\over \beta}\cos(k\pi)$.
Thus the integral is simplified to
$$I=-{i\over 8\pi}
{1\over R A_\beta}
\sum_{k=-\infty}^{k=+\infty}
e^{-i{(E-E_0)\over A_\beta}(2R\sin z_k)}
{1\over   \sin z_k}\
{1\over 1-\beta\omega R\cos z_k }
 \eqno {(27)}$$
 From this expression it is clear that the explicit dependence on $\beta$ is
through the location of the poles.
The contribution from the pole corresponding to $k=0$ is the dominant one
and it depends on $\beta\omega R$ only.

We now show that the summation above is convergent. Using symmetry properties
of the integrand we can see that the residues for $k<0$ are the negative
complex conjugates of the residues for $k>0$. Hence the convergence of the
series is determined by the convergence of
$$-{i\over 8\pi}
{1\over R A_\beta}
\sum_{k=0}^{k=+\infty}
{1\over   \sin z_k}\
{1\over 1-\beta\omega R\cos z_k }
 \eqno {(28)}$$
for large $y$. We can take $x=\pi/2$, $\sinh y=-e^{-y}/2$,
$\cosh y=e^{-y}/2$, and $e^y={\beta\omega R\over \pi-2\beta k\pi}$.
Thus by comparison with the $\sum{1\over k^2}$, it can be seen that the
series is convergent.

We give below numerical values of the residues for typical values of the
parameters. As a physically realistic case we take
$$\beta=0.9, \quad R=1, \quad \beta\omega R=0.6, \quad
E-E_0=1.$$
Then the contributions from the first few poles are given below.
$$-0.007334,\qquad -0.00010600,\qquad 5.1\times 10^{-7},\qquad
2.59\times 10^{-6},\qquad -6.16\times 10^{-8},\qquad$$
$$~~~~~~~~~~~~~~-3.37\times 10^{-7},\qquad -3.92\times 10^{-8}$$

The contributions from the
residues for large $\Delta \tau$ become quickly comparable with
computational precison and it is not  meaningful to attempt a computation
for large $\Delta \tau$ using residues only.
Hence as an approximation for the integral for infinite time we use a contour
consisting of the union of the original contour with $\Delta \tau < 1000$ and
the horizontal line joining the two end points. The integral for finite but
very large time is obtained as the sum of residues inside the contour minus
the value of the integral along the horizontal line.

We have obtained the plots of these integrals for various combinations of
$\beta$, $\omega$ and $R$ values. By numerical integration one can verify
that the integral converges to a finite value as the range of integration
is increased.  We calculated the value of the  integral for $\omega R =0.6$
and for $ \omega R =0.8 $ for $\beta$ ranging between unity and 0.61.
These values are plotted in Figures 1.a and b.   Here the value of the
integral for $\beta=1$ is subtracted from the value found for a particular
$\beta$. We find that these two figures can be fitted to the function
$$ A_{\omega R}[
\exp ( 14 \pi (1-(\omega R )^2)^{1/3} \omega R (\beta-1)
\sqrt {1-(\omega \beta R)^2})-1] $$
where the $A_{\omega R} $ varies with $\omega R$ as given in Fig.2.
We also calculated the behaviour of the integral for constant
$\beta \omega R$
as $\beta $ ranges from unity to .61.   This behaviour is seen in Fig.3
and can be fitted to the function
$$A_{\beta \omega R} (\exp{ a_{\beta \omega R} (\beta-1)} -1) . $$
Here we find that $a_{\beta \omega R} $ is much smaller
than the constant in the
previous case; its sign is different and is only five percent of
that number in magnitude.

In both cases we conclude that there is a distinct difference when the
cosmic string is present compared to the case when it is absent.
We see that the general behaviour  does not change considerably as time
ranges from small to large values.

\medskip
\noindent
{\bf The case $E-E_0<0$.}
\smallskip
In the previous  calculation we assumed $ E> E_0 $.
If $E<E_0$, still using the change of variables
$t-t'-i\epsilon \to 2R \sin z$,  the integral over the real line is mapped
to the contour determined by the imaginary part of (24), but in this case
we cannot  close the contour as $y\to -\infty$,
as the integrand does not vanish there.
We can however use the following symmetry argument to obtain a
closed contour and use Cauchy's theorem to evaluate the integral.
The contour of integration in the lower half plane can be deformed to a
nearly
rectangular path consisting of the union of the lines
$\{x=\pm \pi/2, y<\epsilon\}$
and $\{y=-\epsilon, -\pi/2<x<\pi/2\}$.
Note that the integrand is invariant under $y\to -y$
when $x=\pm\pi/2$.  We consider the contour consisting of the
union of the lines $\{x=\pm \pi/2, y>-\epsilon\}$, and $\{y=-\epsilon,
-\pi/2<x<\pi/2\}$. The integral over the line segments $\{x=\pm \pi/2,
-\epsilon<y<\epsilon\}$ arise as additional terms in the integral over this
new contour but these terms go to zero as $\epsilon \to 0$ provided that
there are no poles on these lines. The new contour can be closed as
$y\to \infty$ and residue calculus can be used to compute the integral.


We looked  for the poles of the integrand in this region.
Since equations (25) are invariant for -y going to y, we get the same
number of poles as the previous case.  There are extra ones,though.
There is a pole at $z=0$ for all values of $\beta$ including unity. The
existence of extra poles when $\beta$ is less than one depends strongly on
what this value is. For values of $\beta$ close to unity,  we could not find
any new poles. We found the first pole for
$\beta< {{\pi}\over {2(\pi-1)}} $  if we
set $\omega R$ equal to unity. For values of angular velocity less
than unity, we find poles for smaller values of $\beta$. If $ \beta = 1/2 $
a new pair of  poles exist for  any finite value of $ \omega R $.

We checked the  presence   of other poles in the rectangular region by
studying the equations carefully and by performing contour integrations
around finite regions numerically which gave zero  within sensible limits.

When $ \beta > {{\pi} \over { 2(\pi-1)}} $ we have only one extra pole at $z$
equals zero.
We can evaluate the residue  corresponding to this pole. In the presence
of the  string we get
$$ {F_{1} \over {T}}= {{(E_0-E)  \Theta (E_0-E)}  \over {32 \pi R
(1- \omega ^{2} \beta  ^{2} R^{2} ) A^{2}_{\beta}}} .  \eqno {(29)}$$
When there is no string we get the similar expression where $\beta= 1$.
Then   we see that the expressions given for these two cases
are identical if we take only the extra poles into account.

If we have $\beta$ taking values which seems to be excluded by experiments,
however, we get the signature    of the string in the residue of two new
extra poles.

If $1/2 \leq \beta < {{\pi} \over {2(\pi-1)}} $, we may have a new pair of
poles for appropriate values of $ \omega R $. If $\beta = 1/2 $, then any
positive value of $\omega R $ allows one pair of new poles.  For $ \beta=1/2$
and $ 2\omega R \sin {{3\pi} \over {8}} = {{\pi}/2} $,
then we can evaluate the residue.   The extra contribution is given by
$$ {1\over {32 \pi R A_{\beta}}} {\Theta (E_0-E) \over {\sin{3 \pi \over
{8}}}}\sin \left({ (E-E_0)R \sin {3 \pi/4} \over {A_{\beta}}}\right). \eqno
{(30)}$$
In this expression  both the value of $\beta$ and $\omega R$  are fixed by
the equations given above.

If $\beta <1/2 $, a second pair of poles may come up depending on the
value $\omega R $ takes.  If $ \beta =1/3 $, we have the second pair of
poles for any value of $ \omega R $.  Only the value of the residue
depends on $ \omega R $.
Similar behaviour goes on.  For $ 1/4 < \beta < 1/3 $ another pair of poles
is possible. For $ \beta =1/4 $, we have the new pair for any $ \omega R $,
etc.. At the end we got a formula like the one given in reference 8 ,
also given in ref's 4 and 9, for similar processes.
This extra contribution reads
$$ F_{extra}/T = \Theta (E_0-E) \sum_{i=1}^{p-1} {{C_i \over
{2R A_{\beta}}}} \sin \left((E-E_0)R {{\sin {2 \theta_i }} \over {A_{\beta}}}
\right)  \eqno {(31)} $$
where $C_i, \theta_i$  are constants depending on the location of the pole.
Here $ {1 \over {\beta}}=p$
where $p$ is an integer.  If $ {1/ \beta } $ is not an integer
then the sum goes up to the integer less than $1/ \beta$  or $1/{\beta-1}$
depending on the value of $\omega R$.


\medskip
\noindent
{\bf 3.2 Dedector response for finite (small) time.}

In this section we study the behaviour of the integrand
$$I_\beta(\Delta\tau)={1\over (2\pi)^2}{1\over \beta r_1 r_2}
e^{-i\Delta E \Delta \tau}
{1-\Delta^{2/\beta} \over 1+\Delta^{2/\beta}-2\Delta^{1/\beta}\cos
(\Delta\phi)}\eqno{(32)}$$
near $\Delta \tau=0$.

Let $p=\Delta t- i\epsilon$. We shall express the integrand in terms of $p$
and obtain  first three terms of its Laurent expansion around $p=0$. At the
coincidence limits $\Delta z=0$, $\Delta r=0$, we have
$$r_1=ip,\quad \quad r_2=\sqrt{4R^2-p^2},\quad\quad
\Delta \phi=\omega (p+i\epsilon),\quad \quad \Delta\tau={1\over A_\beta}
(p+i\epsilon).\eqno{(33)}$$
Inserting these in the integrand we obtain an expression $I_\beta(p)$.
We obtain the Laurent expansion of $I_\beta(p)$ using REDUCE as follows.
$$\eqalignno{
I_\beta(p)=&{e^{\Delta E\epsilon/A_\beta}\over
               4\pi^2 (\beta^2\omega^2 R^2 -1)}{1\over p^2}
          -i{e^{\Delta E \epsilon/A_\beta}\over
               4\pi^2 A_\beta (\beta^2\omega^2 R^2 -1)}{1\over p}\cr
          &+{e^{\Delta E \epsilon/A_\beta}\over
            48 \pi^2 \beta^2 R^2}
            \left[ (\beta\omega R)^4 +2 \beta^2 (\beta\omega R)^2
               -2  (\beta\omega R)^2 -\beta^2 +1\right]
          +{e^{\Delta E \epsilon/A_\beta}\over
            48 \pi^2 A_\beta^2 }
            \left[ \Delta E^2  (1-(\beta\omega R)^2\right]\cr
          &+\dots&(34)\cr}$$

Note that in the limit $\epsilon \to 0$, $p$ is proportional to $\Delta \tau$
and the ${1\over p} $ term has no contribution when integrated over a
symmetric interval. Thus the divergence is due to the ${1\over p^2}$ term.

Now let  $\beta\omega R$ be fixed and consider the difference
$ I_\beta(p)-I_{1/k}(p)$.
In this case, as
the coefficients of ${1\over p^2}$ and ${1\over p}$ do not depend on $\beta$
the divergences are eliminated. Then we have
$$ I_\beta(p)-I_{1/k}(p)=
          e^{\Delta E \epsilon/A_\beta}
          {1-\beta^2 k^2 \over
            48 \pi^2 \beta^2 R^2 }+O(p).\eqno{(35)}$$

Thus the difference
$ I_\beta(p)-I_{1/k}(p)$ with $\beta\omega R$ fixed is regular
at $\Delta\tau=0$ and there is a qualitative difference between the cases
$\beta=1/k$ and $\beta\ne 1/k$.

Repeating a similar Laurent  expansion for a stationary detector, it can be
seen that the difference
$ I_\beta(p)-I_{1/k}(p)$ is nonzero for $\beta\ne 1/k$. This result agrees
with  Reference 9, where there is nontrivial detector response for a
stationary   detector when the interaction time is finite.

The method of obtaining a Laurent expansion of the integrand around a pole can
be applied without specifying the trajectory explicitly.
We briefly outline the method but omit explicit calculations.
We assume that $r_1$, $r_2$, $\Delta\tau$ and $\Delta \phi$
are certain analytic functions of $p$ such that
$$
\lim_{p\to 0}r_1=0,\quad \lim_{ p\to 0}r_2\ne 0,\quad
\lim_{ p\to 0}\phi=O(\epsilon)\quad \lim_{p\to 0} \Delta \tau=O(\epsilon).$$
We can then obtain the Laurent expansion of $I_\beta(p)$ with
straightforward but messy computations.
Under these general assumptions, the Laurent expansion starts with
$p^{-2}$ term and
with additional symmetry assumptions it is possible to ensure that
$p^{-1}$ term has no contribution if integrated over a
symmetrical interval.
The integrand is regularized by subtracting the ``background" contribution
and the constant term  in the regularized integrand in proportional to
$1-\beta^2k^2$ as before.

 The ``background" for each trajectory has to taken a spacetime
with $\beta=1$ with trajectory parameters such that the expressions involved
in the divergent terms are kept fixed. For example, in the case of a detector
in a spacetime with $\beta=0.9$, $R=1$, $\omega=0.6$, so that
$\beta \omega R=0.54$, the ``background" has to be
a detector moving in  the Minkowski space with say $R=1$, $\omega=0.54$.

\bigskip
\noindent
CONCLUSION

Here we first reviewed the already known material concerning whether the
presence of a cosmic string can be felt by an Unruh detector, in a new
formalism. We ,then extended our calculations to the undecided case when a
detector revolves around a string with a constant angular velocity. This
case was first studied in reference 8 and a definite answer was not obtained.
We found that both the integrand and the resulting integral are different
from the expressions we get in the absence of the cosmic string.
We find a qualitative difference when we study the integrands, though.
For $\beta \neq 1/k$, the subtracted expression does not vanish as the
argument approaches zero, whereas the contrary is true when $\beta=1/k$.
This effect can be seen from the integrand expression for the detector by
performing the integration over a small interval, and dividing the integral
by the interval.  In the limit the interval goes to zero, we will get zero
if $ \beta=1/k$ and a non zero result if $\beta \neq 1/k$.

If we study the case when $E_0$ is greater then $E$, which describes the
deexcitation of the detector, we find extra poles if $\beta$ is less than a
certain value.
For the critical value of $ \beta< {{\pi} \over {2(\pi-1)}}$,
there is a new contribution which occurs only
for very fast particles with velocities close to that of light .  When
$\beta=1/2$, a particle with any finite velocity will sense this effect.
For even smaller values of $\beta $ we have additional
contributions which first occur for fast or slow particles depending upon the
value of $\beta$.

A quantization condition is seen to set in for $\beta = 1/ {p} $ , where $p$
is an integer. An extra contribution is possible only if we pass a new
threshold.
To be certain of the new contribution, for any non zero value of
$ \omega R $, we must go to the next value, $\beta = {1 \over {p+1}} $.
This behaviour reminds us of the Bohr-Sommerfeld
condition of fitting an integer number of waves on the cone, this number
depending on $p$.  If we are between  $p$ and $p+1$  we may be able to fit
another one depending  on the value of $\omega R $ which decides where
on the cone this wave will be located.

We think that this quantization phenomena depending on the value of $\beta$
and the ``new phenomena " that occurs at a critical value of $\beta $ where
we first have a new pair of poles should be investigated in other physical
processes as well.
\bigskip
{\bf Acknowledgement.}  We would like to thank  Yavuz Nutku and A.N. Aliev
for discussions, H. Kaya, C. \" Ozben and A.T. Giz for assistance in
numerical calculations. This work is partially supported by TUBITAK,
the Scientific and Technical Research Council of Turkey.  M.H. is also
supported by TUBA, Academy of Sciences of Turkey.
\vfill\eject
\noindent {\bf {REFERENCES} }

\item {1.}   {\it { The Formation and Evolution of Cosmic Strings}}, ed. by
G.Gibbons, S. W. Hawking and T. Vachaspati,Cambridge Univ. Press (1990) .

\item {2.} A. Vilenkin and E.P.S. Shellard, {\it Cosmic Strings and Other
Topological Defects}, Cambridge University Press, Cambridge (1994).

\item {3.} M.B.Hindmarsh and T.W.B. Kibble, Reports on Progress in Physics,
58(1995)477.

\item {4.}  B.F. Svaiter and N.F.Svaiter, Class. and Quant. Grav. 11
(1994) 3471.

\item {5.}  L. Iliadakis, V. Jasper and J. Audretsch, Phys. Rev. D51 (1995)
2591.

\item {6.}  W. Unruh, Phys. Rev. D14 (1976) 870.

\item {7.}  B.DeWitt, (in {\it {General Relativity}}, eds. S.W.Hawking and
W.Israel, Cambridge University Press) 1979.

\item {8.} P.C.W. Davies and V. Sahni, Class. Quant. Grav. 5 (1988) 1.

\item {9.} V.Frolov, V.D. Skarzhinsky and A.M. Amirkhanjan, Astroparticle
Physics J. 3 (1995) 197.

\item {10.}  N.D.Birrell and P.C.W. Davies {\it { Quantum Fields in Curved
Space}}, Cambridge University Press (1982).

\item {11.}  A.G.Smith. Tufts preprint (1986) Published in our Reference 1,
p. 263-292; \hfill\break
T.M.Helliwell and D.A.Kon\-kows\-ki, Phys.Rev. D34 (1986) 1908;
B.Linet, Phys. Rev. D33 (1986) 1833, Phys. Rev. D35 (1987) 536.
\vfill\eject

\centerline{FIGURES}
\vskip 3truecm
\baselineskip=18pt
\item{Fig 1a.} Numerical integration results for the detector response
function when $\beta$ varies from 0.61 to 1 and $w R=0.6$.

\item{Fig 1b.} Numerical integration results for the detector response
function when $\beta$ varies from 0.61 to 1 and $w R=0.8$.

\item{Fig 2.} The variation of the coefficient as a function of $w R$.

\item{Fig 3.} Numerical integration results for the detector response
function when $\beta$ varies from 0.61 to 1 and $w\beta R=0.6$.

\end